\def \pg{PG 1553+113}

\documentclass[]{aa}  

\usepackage{graphicx}
\usepackage{hyperref}
\usepackage{txfonts}
\usepackage{xcolor}
\usepackage{ulem}
\begin{document}

   \title{Unveiling the periodic variability patterns of the X-ray emission from the blazar \pg}

   \subtitle{}

   \author{T. Aniello
          \inst{1,2,3}
          \and L. A. Antonelli
          \inst{1}
          \and F. Tombesi
          \inst{1,3,4,5,6}
          \and A. Lamastra
          \inst{1}
          \and R. Middei
          \inst{1,7}
          \and M. Perri
          \inst{1,7}
          \and F. G. Saturni
          \inst{1,7}
          \and A. Stamerra
          \inst{1,8}
          \and F. Verrecchia
          \inst{1,7}
          }

   \institute{INAF -- Osservatorio Astronomico di Roma,
              Via Frascati 33, I-00078 Monte Porzio Catone (RM), Italy\
         \and
             ``Sapienza'' Universit{\`a} di Roma -- Dip. di Fisica, P.le A. Moro 5, I-00185 Roma, Italy\
         \and
             Universit{\`a} degli Studi di Roma ``Tor Vergata'' -- Dip. di Fisica, Via della Ricerca Scientifica 1, I-00133 Roma, Italy
         \and
             INFN -- Roma Tor Vergata, Via della Ricerca Scientifica 1, I-00133 Roma, Italy 
         \and
             University of Maryland -- Dept. of Astronomy, College Park, MD-20742, USA
         \and
             NASA -- Goddard Space Flight Center, Greenbelt Rd. 8800, MD-20771 Greenbelt, USA
         \and
             ASI -- Space Science Data Center, Via del Politecnico snc, I-00133 Roma, Italy
         \and
             Scuola Normale Superiore di Pisa, P.zza dei Cavalieri 7, I-56126 Pisa, Italy
             }

\date{Received 18 March 2024; accepted 5 April 2024}


\abstract
  {The search for periodicity in the multi-wavelength high variable emission of blazars
  is a key feature to understand dynamical processes at work in this class of active galactic nuclei. The blazar \pg\ is an attractive target due to the evidence of periodic oscillations observed at different wavelengths, with a solid proof of a 2.2-year modulation detected in the $\gamma$-ray, UV and optical bands.
  We aim at investigating the variability pattern of the \pg\ X-ray emission using a more than 10-years long light curve, in order to robustly assess the presence or lack of a periodic behavior whose evidence is only marginal so far.
  We conducted detailed statistical analyses, studying in particular the variability properties of the X-ray emission of \pg\ by computing the Lomb-Scargle periodograms, which are suited for the analyses of unevenly sampled time series, and adopting epoch folding techniques.
  We find out a modulation pattern in the X-ray light curve of PG 1553+113 with a period  of $\sim$1.4 years,
 about 35\% shorter than the one observed in the $\gamma$-ray domain. Our finding is in agreement with the recent spectro-polarimetric analyses and supports the presence of more dynamical phenomena simultaneously at work in the central engine of this quasar.}

   
   \maketitle
   
%

\section{Introduction}

\color{black}

    Blazars are a class of active galactic nuclei (AGN) characterized by extreme luminosity and variability over the entire electromagnetic spectrum. Their emission is dominated by a single component, i.e. a jet of relativistic particles directly pointing towards the observer \citep{Urr95}. From a spectroscopic point of view, these sources show a peculiar spectral energy distribution (SED) that is characterized by a double-humped shape. The low-frequency peak, that can be observed from the radio up to the X-ray domain, is attributed to synchrotron radiation arising from high-energy electrons that spiral around magnetic field lines \citep{Pad95}; its properties mainly depend on the strength of the magnetic field and on the energy distribution of the relativistic electrons in the jet \citep{Mar92}. The second hump is usually observed in the X-to-$\gamma$-ray range, and its origin is commonly associated with inverse Compton (IC) emission \citep{Ghis92,Abd11,Zdz15}, or synchotron self-Compton where the seed photons emerge from either external radiation fields (external Compton; EC) or the synchrotron radiation of the jet itself, respectively.  Several evidences, from SED modeling \citep[e.g.,][]{Abdo2011-II} and energetic considerations \citep{Zdz15,Liodakis2020-II} to observations of correlated flux variations across different wavebands \citep[e.g.,][]{Agudo2011,Agudo2011bis,Liodakis2018}, and, recently,  polarimetry \citep{Middei2023,Peirson2023}, support the leptonic origin of the second hump of the blazars SED.

    \begin{figure*}[htbp]
   \centering
   \includegraphics[width=\textwidth]{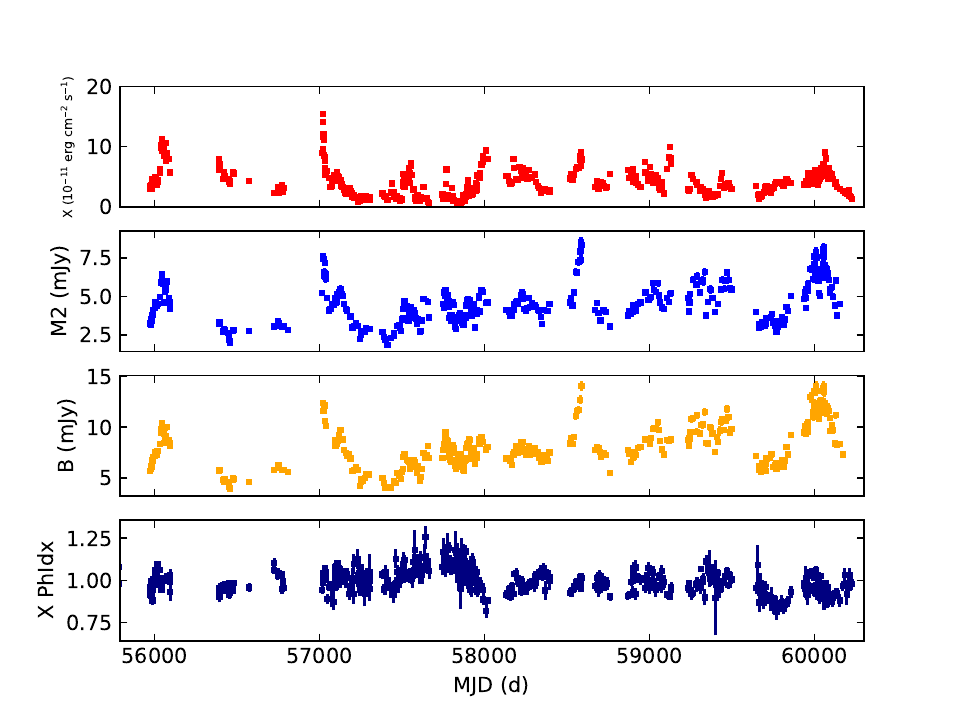}
      \caption{MWL LCs of PG1153+113 in B (yellow dots), M2 (blue dots), X-ray (red dots) bands and spectral index (dark blue dots) from 2012 to 2023 of Swift satellite. The X-ray fluxes are in 0.2-10 keV band. B, M2 and X-rays LCs are showed along with the corresponding 1$\sigma$ uncertainties.}
        \label{fig:mwllc}
   \end{figure*}
    
    A hallmark of the blazar phenomenon is the prominent flux and spectral variability, which is thought to arise from changes in the jet properties such its particle density, magnetic field strength and orientation \citep{Pad95}. Variability can be observed over different timescales from hours up to decades \citep{Kel92}. In this context, the search of periodic signals is object of growing attention, and the high synchrotron peaked source PG 1553+113 represents one of the most interesting cases. \pg, a blazar with optical magnitude $V\,\sim\,14.5$ at redshift $z\,\sim\,0.4 \div 0.5$ \citep{2010ApJ...720..976D}, shows evidence of a periodicity ($T \sim 2.2$ yr) in the $\gamma$-ray band ($E \geq\,100\,\rm MeV$) sampled by the {\it Fermi}-LAT satellite and at lower frequencies  \citep[$R$-band;][]{2015ApJ...813L..41A,2017MNRAS.465..161S,Pen23}. A possible explanation for the periodic signal was proposed by \citet{2017MNRAS.465..161S}, who discussed a scenario where a couple of asymmetric super massive black holes (SMBHs), of which the smallest carries a jet, interacts producing a precession of the jet itself. 

    The temporal properties of the PG 1553+113 X-ray emission are instead more debated: \citet{Hua21} claimed the finding of the same periodicity with respect to the $\gamma$-ray emission within a scenario in which both SMBHs possess a jet. Conversely, \citet{Pen23}, working on a blazar sample, identified a periodicity of $\sim$1.5 years with a significance of $\sim$2$\sigma$. In this paper, we report on the temporal properties of PG 1553+113 by investigating the data collected in the rich {\it Swift} archive, which provides 617 X-ray and $>$400 optical/UV observations. By constructing the source multi-wavelength (MWL) light curves (LCs), we correlate the various emission bands and search for an unambiguous periodic signal at each wavelength with robust methods of time series analysis, such as the construction of the Lomb-Scargle periodograms \citep{1976Ap&SS..39..447L,1982ApJ...263..835S} and the application of epoch folding techniques \citep[e.g.,][]{lar96}, with a particular attention to the X-ray signal in order to identify with enough significance an associated periodicity.

    The paper is organised as follows: in Sect. \ref{sec:obsred} we illustrate the data selection and reduction; in Sect. \ref{sec:corr} we compute correlations between different bands; in Sect. \ref{sec:xvar} we present the X-ray variability analysis; in Sect. \ref{sec:disc} we discuss our findings; finally, in Sect. \ref{sec:sum} we summarize the obtained results. Throughout the text, we adopt a concordance cosmology with $H_0=70$ km s$^{-1}$ Mpc$^{-1}$, $\Omega_M=0.3$ and $\Omega_\Lambda=0.7$.

    \section{Observations and data reduction}
        \label{sec:obsred}
        
    Launched in 2004, the {\it Swift} satellite carries the X-ray Telescope (XRT), which is sensitive in the $0.2 \div 10$ keV energy band, and the UV/Optical Telescope (UVOT), capable of observations in the $170 \div 600$ nm band. XRT has high spatial resolution of 18$''$, allowing for precise localization of celestial objects. XRT and UVOT operate simultaneously, to enable concurrent observations in different electromagnetic bands \citep{Rom05,Bur05}. Since our e retrieved the X-ray, UV and optical reduced data from 2005 to 2023 from the {\it Swift} public mirror archive \footnote{Available at \url{https://swift.ssdc.asi.it/}.} of the Space Science Data Center (SSDC) at the Italian Space Agency (ASI). In this analysis, we excluded the data taken before 2012, characterized by a sparse temporal sampling, to avoid the introduction of biases that could be ascribed to time intervals containing no data points.

        \begin{table}[htbp]
        \centering
        \resizebox{\columnwidth}{!}{
        \begin{tabular}{lcc}
        \hline
        \hline
        \multicolumn{3}{l}{ }\\
             Correlation &  Pearson coeff.  &  Degrees of freedom  \\
             \multicolumn{3}{l}{ }\\
             \hline
             \multicolumn{3}{l}{ }\\
           U/X  &  0.54  &  265  \\
           B/X  &  0.50  &  264  \\
           V/X  &  0.48  &  254  \\
           W1/X &  0.57  &  279  \\
           M2/X &  0.58  &  269  \\
           W2/X &  0.60  &  278  \\
           X PhIdx/flux &  0.55  &  303  \\
           \multicolumn{3}{l}{ }\\
           \hline
        \end{tabular}
        }
         \caption{Results of the correlation analysis (Pearson correlation coefficients and number of degrees of freedom) between pairs of wavebands, and between the X-ray photon index and flux.}
        \label{tab:tab1}
    \end{table}
    
     

The {\it Swift}-XRT observations were carried out in the Windowed Timing (WT) and Photon Counting (PC) readout modes. The data were first reprocessed locally with the {\footnotesize XRTDAS} software package (version {\ttfamily v3.7.0}), developed by the ASI-SSDC and included in the NASA-HEASARC {\footnotesize HEASoft} package\footnote{Available at \url{https://heasarc.gsfc.nasa.gov/docs/software/heasoft/}.} (version {\ttfamily v6.31.1}). Standard calibration and filtering processing steps were applied to the data using the xrtpipeline task. The calibration files available from the {\itshape Swift}-XRT {\footnotesize CALDB} (version {\ttfamily 20220803}) were used. Events for the temporal and spectral analysis were selected within a circle of 20-pixel ($\sim$47$''$) radius, while the background was estimated from nearby circular regions with a radius of 40 pixels. For each observation, the X-ray energy spectrum was first binned with the {\ttfamily grppha} tool of the {\footnotesize FTOOLS} package\footnote{Available at \url{https://heasarc.gsfc.nasa.gov/ftools/}.} to ensure a minimum of 20 counts per bin, and then modeled using the {\footnotesize XSPEC} software package\footnote{Available at \url{https://heasarc.gsfc.nasa.gov/xanadu/xspec/}.} adopting a single power-law model.

    We obtained dereddened UV and optical fluxes with a dedicated ASI-SSDC pipeline for the analysis of UVOT sky images \citep[][]{Gio2012}. We first executed the aperture photometry task included in the UVOT official software from the {\footnotesize HEASoft} package (version {\ttfamily v6.26}), extracting source counts within a standard 5$\arcsec$ circular aperture and the background counts from three circular 18$\arcsec$ apertures that were selected to exclude nearby stars. We then derived the source dereddened fluxes by applying the official UVOT calibrations from the {\footnotesize CALDB} \citep[][]{brev11}, and adopting a standard UV/optical mean interstellar extinction law \citep[][]{Fitzpatrick1999} with a mean $E(B-V)$ value of 0.0447 mag \citep[][]{Schlafly11}. We show the X-ray, UV (M2 filter) and optical ($B$-band) LCs obtained in this way in Fig. \ref{fig:mwllc}, along with the X-ray photon index: a visual inspection already reveals considerable flux variability. Furthermore, while the UV and optical trends can be overlapped, the X-ray band exhibits some peaks that do not appear in the other bands, hinting to a possible X-ray periodicity that is different with respect to the optical/UV one.

   
    
    
    \section{Correlations between different bands, and between photon index and X-ray flux}
    \label{sec:corr}
    
    %

    \begin{figure}[htbp]
   \centering
   \includegraphics[width=\columnwidth]{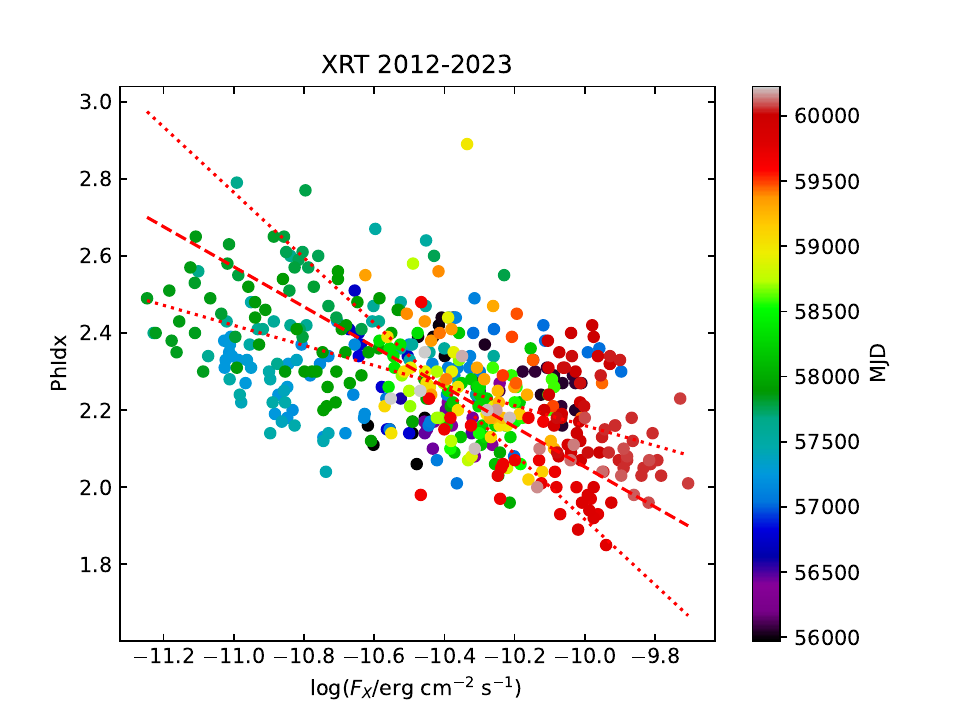}
      \caption{\pg\ X-ray photon index–to-flux correlation. The two linear fits of photon index versus flux and vice-versa ({\it red dotted lines}) are shown along with the corresponding bisector fit ({\it red dashed line}). The data points are color-coded on the basis of their MJD.}
         \label{fig:FigPhIdFluxRatio}
   \end{figure}

    We first proceeded studying the correlations among the various bands, and between X-ray photon index and flux. The photon index and the X-ray flux of \pg\ in the 0.3-10 keV band are moderately anti-correlated (see Fig. \ref{fig:FigPhIdFluxRatio}) as we computed a Pearson coefficient \citep{bev03} of $r = -0.55$ and an associated null-hypothesis probability $p($>$r) = 3.6 \times 10^{-24}$ (see Tab. \ref{tab:tab1}). The X-ray photon index is flatter as the source flux increases. This is commonly observed in blazars \citep[the so-called ``harder when brighter'' behavior; e.g.,][]{Gio21}, and is expected when particles are injected and accelerated in the jet \citep{2010yCat..21880405A}. 
    
\begin{figure*}[htbp]
\centering
\begin{minipage}{0.49\textwidth}
    \includegraphics[scale=0.6]{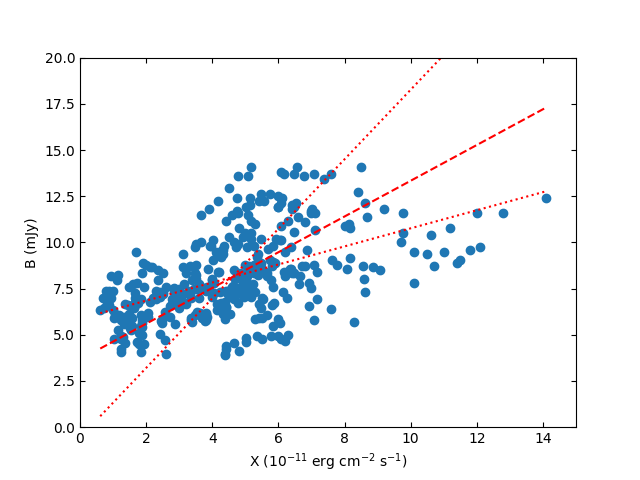}
\end{minipage}
\begin{minipage}{0.49\textwidth}
    \includegraphics[scale=0.6]{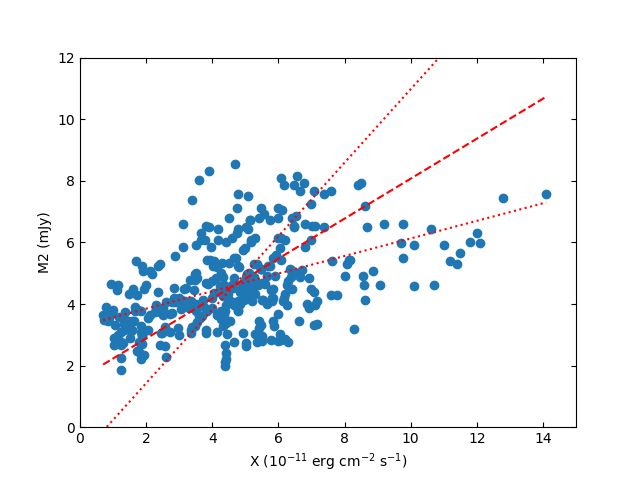}
\end{minipage}
\caption{Correlations of the UV-to-X-ray and optical-to-X-ray bands of \pg. As in Fig. \ref{fig:FigPhIdFluxRatio}, the two linear correlation fits ({\it red dotted lines}) are shown along with the corresponding bisector fit ({\it red dashed line}).}
\label{fig:images}
\end{figure*}

    We then tested the correlation properties between the UV, optical and X-ray bands, respectively. Moderate correlations were found as shown in Fig. \ref{fig:images} and Tab. \ref{tab:tab1}. Such a study highlights the typical behavior of the radiation emitted from blazars in adjacent wavebands such as the X-rays and the optical/UV, since the respective emitting regions are partially overlapping in the jet and produce photons through the same underlying physical process (i.e. the synchrotron or IC radiation) -- or by processes that make it varying in a quasi-simultaneous way \citep{Dhi21}.

    \section{Multi-wavelength variability analysis}
        \label{sec:xvar}
    To investigate the periodic behavior of the X-ray emission in \pg, we performed an analysis of the LC through the employment of numerical methods that are commonly used in studies of time series. In particular, we mainly relied on the construction of the Lomb-Scargle (LS) periodogram \citep{1976Ap&SS..39..447L,1982ApJ...263..835S}, a technique that is widely adopted to search for periodic signals in astronomical data sets \citep[e.g.,][]{Van18,Vio13} and especially suited for analysing unevenly sampled time series \citep{2008MNRAS.385.1279B}. This method consists in the calculation of the power spectral density (PSD) of a time series, estimating the signal likelihood at each frequency on the basis of a least-squares fit of a sinusoidal model to the data.
    For our analysis, we adopted the LS routines contained inside the {\ttfamily AstroPy (v5.0)} Python package \citep{astropy22}.

\begin{figure*}[htbp]
\centering
\begin{minipage}{0.48\textwidth}
    \centering
    \includegraphics[width=\textwidth]{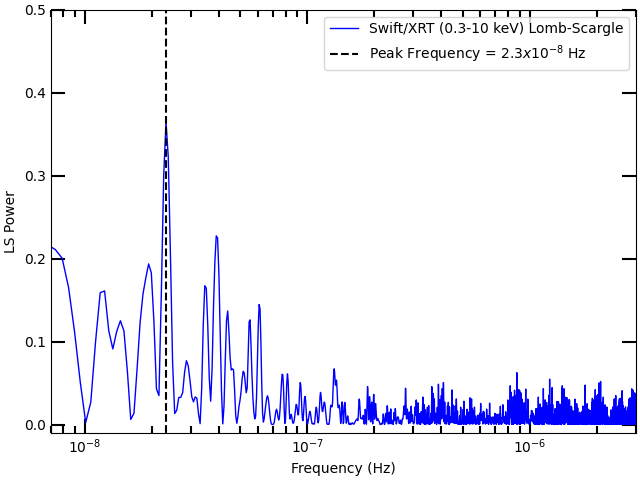}
\end{minipage}
\hfill
\begin{minipage}{0.48\textwidth}
    \centering
    \includegraphics[width=\textwidth]{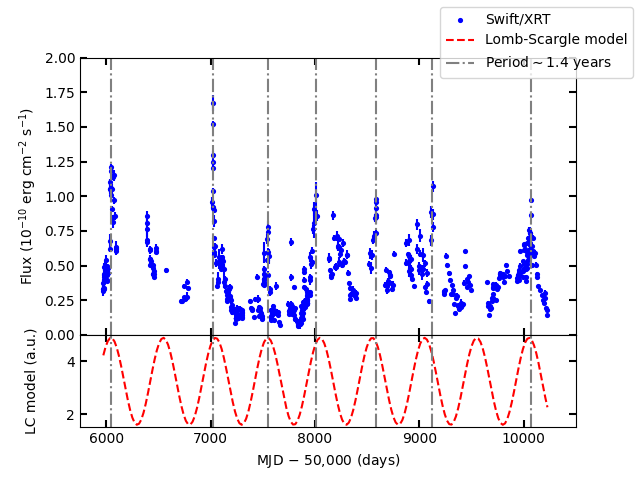}
\end{minipage}
\medskip
\begin{minipage}{0.48\textwidth}
    \centering
    \includegraphics[width=\textwidth]{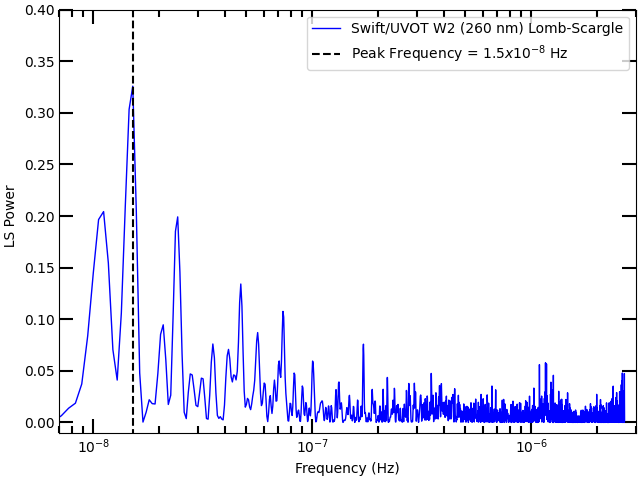}
\end{minipage}
\hfill
\begin{minipage}{0.47\textwidth}
    \centering
    \includegraphics[width=\textwidth]{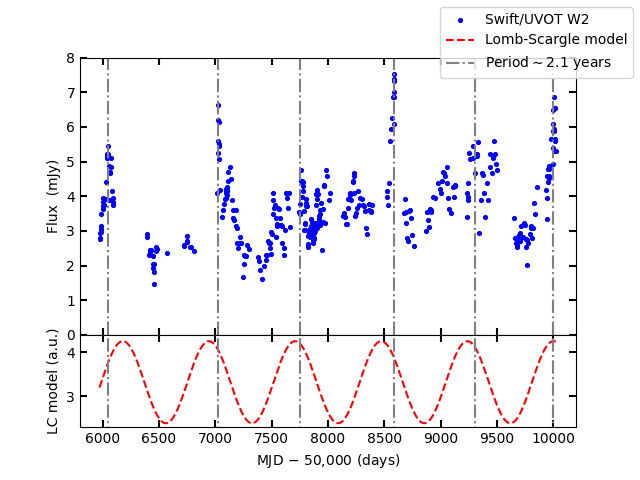}
\end{minipage}
\medskip
\begin{minipage}{0.48\textwidth}
    \centering
    \includegraphics[width=\textwidth]{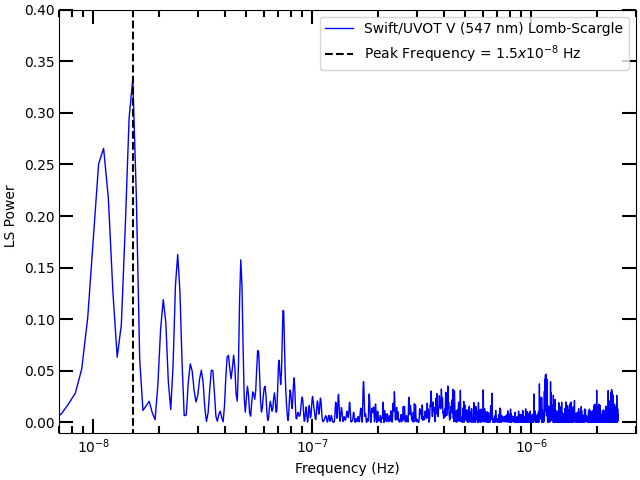}
\end{minipage}
\hfill
\begin{minipage}{0.47\textwidth}
    \centering
    \includegraphics[width=\textwidth]{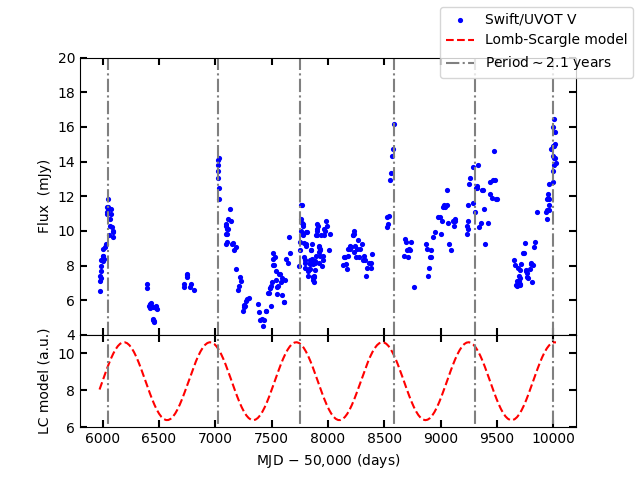}
\end{minipage}
\caption{{\it Left panels:} \pg\ LS periodograms of the X-ray, UV (W2) and optical bands ($V$; {\it blue solid lines}). In each panel, the frequency of the main peak ({\it black dashed line}) is highlighted, and its value is reported ({\it see legend}). {\it Right panels:} X-ray, UV and optical LCs ({\it blue points}), along with the relative sinusoids of periods corresponding to the LS most significant frequencies ({\it red dashed lines}). In such panels, the sinusoid maxima approximately coinciding with flux peaks in the LCs ({\it grey dot-dashed lines}) are marked, and the corresponding period is reported ({\it see legend}).}
\label{fig:images2}
\end{figure*}

    In carrying on our analysis, we already knew that the optical (and the associated UV) LCs exhibit a $\sim$2.2-yr period \citep{2017MNRAS.465..161S}; therefore, we expected that the calculation of the LS periodogram for such LCs should yield a comparable result as a confirmation of the goodness of the method. We thus computed the LS periodograms associated with the \pg\ X-ray, UV and optical LCs, and identified prominent peaks in the PSD. We show the results in the left panels of Fig. \ref{fig:images2}: a visual inspection reveals that, while the main peak associated with the optical and UV variability patterns is located at the same frequency, a clearly prominent peak also appears for the X-ray signal, but -- at variance with the optical/UV case -- is shifted at a larger frequency (corresponding to a shorter period).
    
    We found that such a peak was located at a frequency of $\sim$2.3 $\times 10^{-8}$ Hz, corresponding to $T_{\rm X} \sim 1.4$ years. This value is significantly lower of $\sim$35\% than the $\gamma$-ray period $T_\gamma \sim 2.2$ years identified in the {\it Fermi}-LAT data by \citet{2015ApJ...813L..41A} and in the optical LC by \citet{2017MNRAS.465..161S}, thus hinting at the presence of a different periodic process. To quantify the significance of this peak, we then estimated its false alarm probability (FAP) level, i.e. the probability of accidentally obtaining a given peak power due to noise fluctuations. We considered statistically significant only those peaks whose FAP level was falling below $\sim$10\% \citep{Stu10}, setting the corresponding LS power of $\sim$0.04 to be our 1$\sigma$ significance; in doing so, we obtained a significance of 9.2$\sigma$ for the 1.4-yr X-ray peak. We also repeated the procedure on the entire \pg\ X-ray LC from 2005 to 2023, i.e. including those data that were discarded in the selection described in Sect. \ref{sec:obsred}, to check the persistence of the peak in the LS periodogram: the test yielded the same $T_{\rm X}$ with a significance of 7.5$\sigma$. We argue that this lower significance is due to the presence of large time gaps in the complete X-ray LC (see Fig. \ref{fig:mwllc}).
    
    In the UV and optical LS periodograms, we found a frequency of the most significant peak of $\sim$1.5 $\times 10^{-8}$ Hz that corresponds to $T_{\rm opt} \sim 2.1$ years with a significance of 5.8$\sigma$ (see Fig. \ref{fig:images2}). For completeness, we also tested the LS analysis on the publicly available {\it Fermi}-LAT $\gamma$-ray data\footnote{Available at \url{https://fermi.gsfc.nasa.gov/ssc/data/access/}.} from 2008 to 2023: in doing so, we again obtained the peak at a frequency of $\sim$1.5 $\times 10^{-8}$ Hz with a significance of 7.3$\sigma$, corresponding to the well-known $T_\gamma \sim 2.2$ years found by \citet{2015ApJ...813L..41A}. In light of having retrieved a comparable result with that reported in the literature for the \pg\ joint optical/$\gamma$-ray signal \citep{2017MNRAS.465..161S}, this analysis strengthens our finding of a discrepant $T_{\rm X}$ with respect to the variability period of the LCs at other wavebands.

\begin{figure*}[htbp]
\centering
\begin{minipage}{0.32\textwidth}
    \centering
    \includegraphics[width=\textwidth]{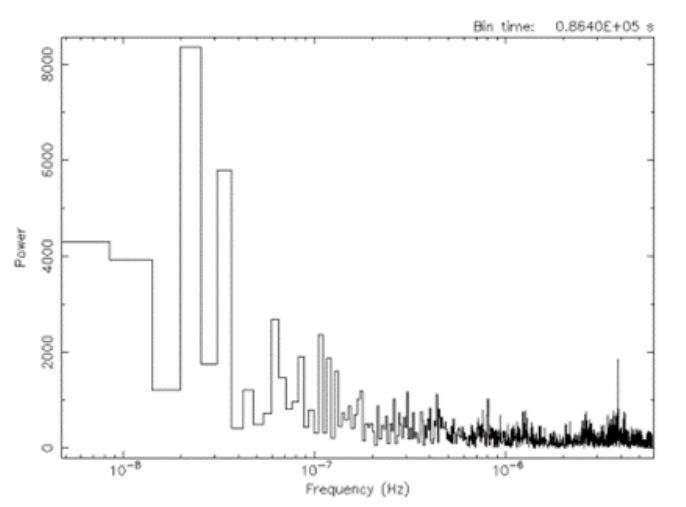}
\end{minipage}
\hfill
\begin{minipage}{0.32\textwidth}
    \centering
    \includegraphics[width=\textwidth]{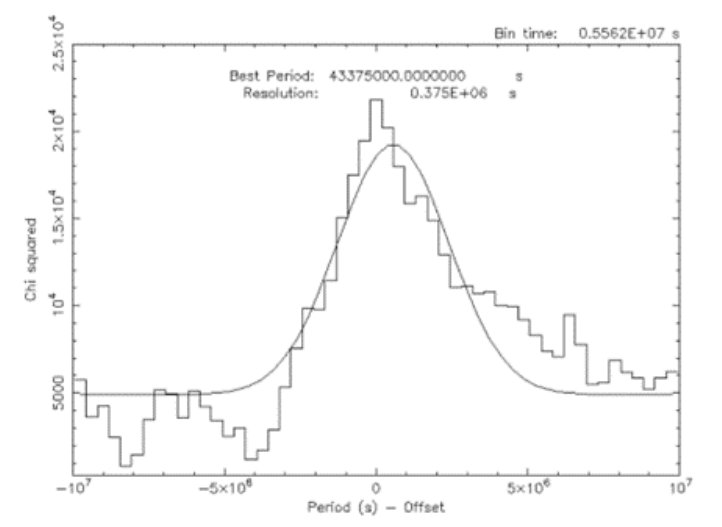}
\end{minipage}
\hfill
\begin{minipage}{0.32\textwidth}
    \centering
    \includegraphics[width=\textwidth]{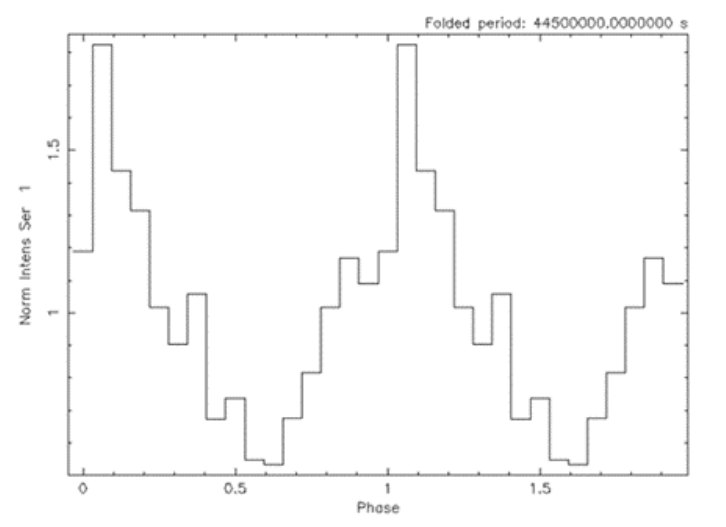}
\end{minipage}
\caption{{\it Left panel:} \pg\ PSD of the X-ray LC obtained with the {\ttfamily powerspec} package of {\ttfamily Xronos}. {\it Middle panel:} best-fit period ({\it solid line}) of the \pg\ X-ray LC obtained with the {\ttfamily efsearch} task, along with its numerical value in seconds corresponding to $\sim$1.4 years ({\it see text}). {\it Right panel:} epoch folding of the \pg\ X-ray LC obtained with the {\ttfamily efold} method on the basis of the best-fit period determined by {\ttfamily efsearch}.}
\label{fig:Xronos}
\end{figure*}

    To further confirm the temporal properties of the \pg\ X-ray emission, we applied two different methods. First, we used the timing tasks provided by the {\ttfamily Xronos} software package\footnote{Available at \url{https://heasarc.gsfc.nasa.gov/xanadu/xronos/xronos.html}.}, designed for the analysis of high-energy astrophysical data \citep{Ste92}. We started from the power spectrum ({\ttfamily powerspec}) method to calculate the PSD of the LCs in each energy band; in this way, we identified the significant peaks corresponding to potential periods in the X-ray LC that match the results obtained with the LS analysis. Then, we applied the epoch-folding search ({\ttfamily efsearch}) method to perform the calculation of the posterior distribution of the LC periodicities: the distribution peak yielded a best-fit period of $\sim$1.4 years, further confirming the LS findings. Finally, we employed the epoch folding ({\ttfamily efold}) method to generate folded LCs at the specific periods of interest. We show the outcome of this cross-check in Fig. \ref{fig:Xronos}.
    
    Then, we also used the ``significance spectrum'' ({\ttfamily SigSpec}) algorithm developed for asteroseismology \citep{Ree07,Chang2011,Mac2014}, which is based on the analysis of frequency- and phase-dependent spectral significance levels $S$ of peaks in a discrete Fourier transform of the signal, through the computation of the probability density function and its associated FAP due to white noise. Like the LS analysis, this method is particularly suited for periodicity studies on sparse data. We thus executed the {\ttfamily SigSpec} algorithm on the UVOT W2 and $V$ time series and finally on the X-ray time series, obtaining the highest significance periods $T_{\rm W2} \sim 2.08$ years ($S_{\rm W2}\sim 28.0$), $T_{V}\sim 2.11$ years ($S_{V}\sim 29.6$) and $P_{\rm X}\sim 1.39$ years ($S_{\rm X}\sim 27.5$), respectively. Also such values are in agreement with those obtained with the LS, thus confirming the goodness of our result.
    
    \section{X-ray period uncertainty estimate and red-noise bias analysis}
        \label{sec:une}
    The calculation of the LS periodogram does not provide any estimate of the uncertainty on the position of the significant peaks. To estimate this quantity, we performed an extensive LS analysis over many altered versions of the \pg\ X-ray LC, in which we replaced each point with a new observing epoch and flux level that differ from the original ones by random amounts extracted from appropriate distributions centered on the actual data. For altering the observing epochs, we adopted uniform distributions, with widths equal to the full width at half maximum (FWHM) of $\sim$1 year derived from the Gaussian fit to the distribution of best-fit periods computed by the {\ttfamily efsearch} task (see Fig. \ref{fig:Xronos}); for the flux values, we instead adopted Gaussian distributions with standard deviations equal to the associated 1$\sigma$ errors. In this way, we produced $10^3$ realizations of the \pg\ X-ray LC; for each realization, we then computed the LS periodogram, and derived the posterior distribution of frequencies of the most significant peak. The statistical analysis of this distribution yielded a best estimate of the \pg\ X-ray period of $T_{\rm X} = 1.41 \pm 0.68$ years.


       Despite the high significance level of our results, we are unable to automatically exclude that the LS analysis is biased by sources of uncertainty that could produce fake signals in the periodogram. This could be due e.g. to non-periodic processes that can mimic a periodic temporal behavior on the time scales of our interest, such as random variability in the AGN flux that is usually distributed according to a red noise spectrum \citep{Bha17,Vau16}. Since the PSD of blazar LCs is of the red-noise type \citep{Vau05}, the power level -- and thus the FAP -- is expected to increase at low frequencies. To assess that our results are not biased by red-noise processes, we simulated the response of the LS analysis to a sample of mock LCs generated according to a pure red noise PSD.
       
\begin{figure}
     \centering
         \includegraphics[scale=0.53]{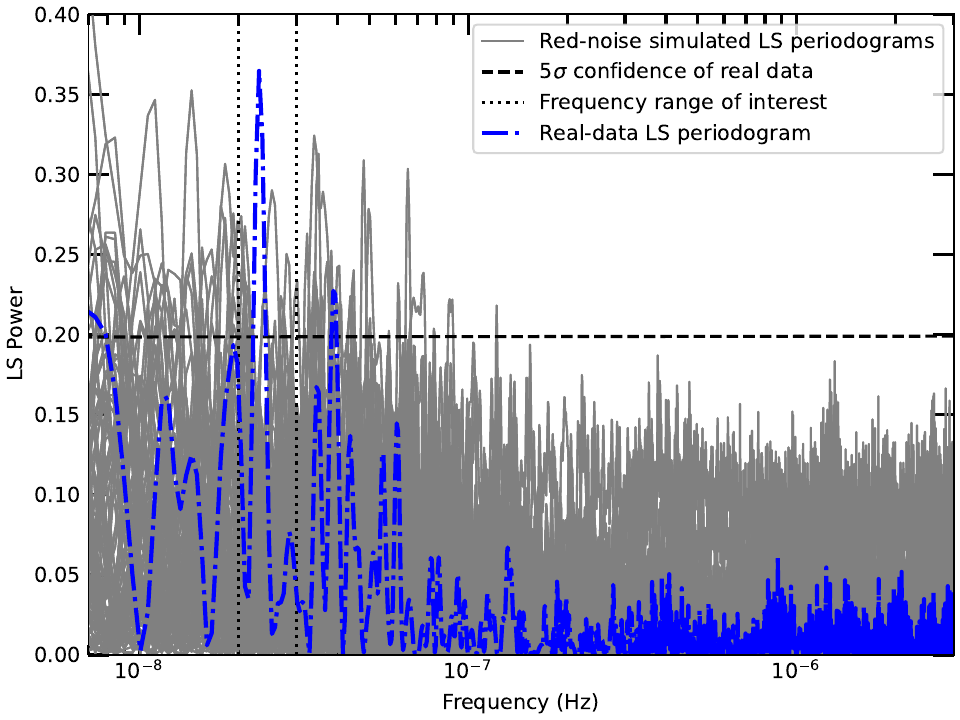}
          \caption{LS periodograms calculated on randomly generated LCs of pure red noise. For plotting purposes, we show $10^2$ ({\it grey solid lines}) out of the total $10^5$ realizations, superimposed to the LS periodogram of the real \pg\ X-ray data ({\it blue dot-dashed line}) and the corresponding 5$\sigma$ significance level ({\it black dashed line}) (see Sect. \ref{sec:xvar}). The relevant frequency interval for our analysis of $(2 \div 3) \times 10^{-8}$ Hz ({\it black dotted lines}) is also highlighted.} 
         \label{fig:LSRedNoise}
       \end{figure}

       To this aim, we generated such LCs by extracting, from a red-noise spectral distribution \citep[e.g.,][]{gar94}, a number of fake flux points corresponding to the amount of X-ray data at our disposal, and associating each of them to an MJD time of our observations. We iterated this process $10^5$ times to reach statistical significance of the results; for each mock LC produced in this way, we computed the associated LS periodogram using the same methodology described in Sect. \ref{sec:xvar}. Requesting a minimum LS power of $\sim$0.04 as our 1$\sigma$ level -- i.e. the same amount associated with the threshold FAP of our real data (see Sect. \ref{sec:xvar}) -- we found that at most $\sim$16$\%$ of our mock LCs rise above 5$\sigma$ in the frequency range of interest $(2 \div 3) \times 10^{-8}$ Hz (see Fig. \ref{fig:LSRedNoise}). Such a fraction is non-negligible, and thus suggests some caution to be adopted in claiming a firm discovery of the $\sim$1.4-year periodicity; nevertheless, these findings point toward the plausible detection of a true periodic signal in the X-ray LC of \pg\ at an $\sim$84\% confidence level.
    
    \section{Discussion}
        \label{sec:disc}
    The presence of multiple periodic patterns in the MWL LCs of \pg\ has been object of various studies \citep{2015ApJ...813L..41A,2017MNRAS.465..161S,Hua21,2022arXiv221101894P,adh23}; recently, there has also been the indication of a potential long-term variability trend in the $\gamma$-ray emission \citep[$\sim$22 years;][]{adh23,Pen23}. Different scenarios have been proposed to explain this behaviour: the most widely accepted one relies upon the presence, inside the \pg\ central engine, of a binary system of SMBHs in which one of the two carries a jet that is gravitationally affected by the SMBH motions \citep{2015ApJ...813L..41A}. This may happen in several ways, such as ({\it i}) a jet precession \citep{2017MNRAS.465..161S}, ({\it ii}) a helical shaping of the jet \citep{2010yCat..21880405A}, or ({\it iii}) instabilities in the jet structure \citep{Hua21}. Alternatively, ({\it iv}) accretion modulations \citep{2018ApJ...854...11T} and ({\it v}) evaporation processes possibly coexisting with disk overdensities \citep{adh23} may also lead to similar results.

    
    The $\gamma$-ray and optical period was already modeled by \citet{2017MNRAS.465..161S}, considering a jet precession with a period $T_\gamma \sim 2.2$ years. To account for an overall different X-ray period, \cite{Hua21} fit the {\it Swift}-XRT LC with a two-jet model, each carried by one of the \pg\ SMBHs, assuming a precession with $T_X \sim 2.2$ years acting on both jets; however, their result is based on the analysis of a less extended X-ray LC with respect to ours. Also \citet{Pen23}, by analyzing the {\it Swift}-XRT data of a sample of $\gamma$-ray detected blazars, found for \pg\ an X-ray period of $\sim$1.5 years with a significance of $\sim$2$\sigma$ after averaging on the temporal properties of the entire sample.

    A possible explanation that does not take into account a binary SMBH system could be the cyclic injection of large quantities of matter from the innermost regions of the central engine into the jet base \citep{Lew19}. If this input of matter occurs regularly, this could produce the emission of a modulated X-ray signal with a different period with respect to that of the $\gamma$-ray, UV and optical emission (due to the jet precession). It is interesting to note that recent observations from the IXPE satellite \citep{Mid23} indicate the presence of different emitting regions in the jet structure, hinting at either a stratified jet or different levels of turbulence inside the jet structure.
        

    The plausible detection of a different time modulation of the \pg\ X-ray emission with respect to the $\gamma$-ray, UV and optical ones has further complicated the road to understand the physical mechanisms acting in the central engine of this blazar. Future MWL observations and studies involving a more detailed modeling of the \pg\ innermost structure (SMBH system, accretion disk, jet, and the respective interplay), will be crucial to eventually explain the origin of its temporal properties.

\section{Summary and future work}
    \label{sec:sum}

    In this study, we have conducted a comprehensive analysis of the X-ray, UV and optical data of the blazar \pg, aimed at investigating the possible presence of a characteristic X-ray periodicity that differs from the already ascertained $\gamma$-ray and optical variability period. We summarize our main findings below:
    
   \begin{enumerate}
      \item the \pg\ X-ray, UV and optical LCs are all moderately correlated to each other according to the Pearson analysis ($r \sim 0.5$); the X-ray photon index is correlated with the X-ray flux in a similar way. This behavior is typical of blazars, where the light is emitted almost entirely from the jet due to the synchrotron and IC processes \citep{Pad95,Mar92}.

      \item The X-ray LC constructed over $\gtrsim$10 observer-frame years of {\it Swift}-XRT data likely ($>$80\% confidence level) exhibits a periodic emission, but with a shorter characteristic period $T_{\rm X} \sim 1.4$ years with respect to that found in the optical and $\gamma$-ray bands \citep[$T_{\rm opt}=T_{\rm UV}=T_{\rm \gamma} \sim 2.2$ years;][]{2015ApJ...813L..41A,2017MNRAS.465..161S,Pen23}.
      
   \end{enumerate}

   Current scenarios are not able to properly explain such a difference in the widely accepted framework of a binary system of SMBHs carrying a preceding jet in the \pg\ central engine \citep{Hua21,2018ApJ...854...11T,2017MNRAS.465..161S,adh23}; therefore, further theoretical investigations and observational data are needed to better disentangle the physical mechanisms that lie at the base of the different variability periods of the \pg\ MWL emission.
    
    
    
 \begin{acknowledgements}
       We thank the anonymous referee for their helpful comments. We acknowledge M. Imbrogno (INAF-OAR) for useful discussion about the use of the {\ttfamily Xronos} software. This article is part of TA's work for the Ph.D. in Astronomy, Astrophysics and Space Science, jointly organized by the ``Sapienza'' University of Rome, University of Rome ``Tor Vergata'' and INAF. Reproduced with permission from Astronomy \& Astrophysics, \textcopyright\ ESO.
\end{acknowledgements}

\color{black}

%
%

\bibliographystyle{aa} 
\bibliography{pg1553.bib} 

\begin{thebibliography}{50}
\expandafter\ifx\csname natexlab\endcsname\relax\def\natexlab#1{#1}\fi

\bibitem[{{Abdo} {et~al.}(2010){Abdo}, {Ackermann}, {Ajello}, {Allafort},
  {Antolini}, {Atwood}, {Axelsson}, {Baldini}, {Ballet}, {Barbiellini},
  {Bastieri}, {Baughman}, {Bechtol}, {Bellazzini}, {Belli}, {Berenji},
  {Bisello}, {Blandford}, {Bloom}, {Bonamente}, {Bonnell}, {Borgland},
  {Bouvier}, {Bregeon}, {Brez}, {Brigida}, {Bruel}, {Burnett}, {Busetto},
  {Buson}, {Caliandro}, {Cameron}, {Campana}, {Canadas}, {Caraveo}, {Carrigan},
  {Casandjian}, {Cavazzuti}, {Ceccanti}, {Cecchi}, {{\c{C}}elik}, {Charles},
  {Chekhtman}, {Cheung}, {Chiang}, {Cillis}, {Ciprini}, {Claus},
  {Cohen-Tanugi}, {Conrad}, {Corbet}, {Davis}, {DeKlotz}, {den Hartog},
  {Dermer}, {de Angelis}, {de Luca}, {de Palma}, {Digel}, {Dormody}, {Silva},
  {Drell}, {Dubois}, {Dumora}, {Fabiani}, {Farnier}, {Favuzzi}, {Fegan},
  {Ferrara}, {Focke}, {Fortin}, {Frailis}, {Fukazawa}, {Funk}, {Fusco},
  {Gargano}, {Gasparrini}, {Gehrels}, {Germani}, {Giavitto}, {Giebels},
  {Giglietto}, {Giommi}, {Giordano}, {Giroletti}, {Glanzman}, {Godfrey},
  {Grenier}, {Grondin}, {Grove}, {Guillemot}, {Guiriec}, {Gustafsson},
  {Hadasch}, {Hanabata}, {Harding}, {Hayashida}, {Hays}, {Healey}, {Hill},
  {Horan}, {Hughes}, {Iafrate}, {J{\'o}hannesson}, {Johnson}, {Johnson},
  {Johnson}, {Johnson}, {Kamae}, {Katagiri}, {Kataoka}, {Kawai}, {Kerr},
  {Kn{\"o}dlseder}, {Kocevski}, {Kuss}, {Lande}, {Landriu}, {Latronico}, {Lee},
  {Lemoine-Goumard}, {Lionetto}, {Llena Garde}, {Longo}, {Loparco}, {Lott},
  {Lovellette}, {Lubrano}, {Madejski}, {Makeev}, {Marangelli}, {Marelli},
  {Massaro}, {Mazziotta}, {McConville}, {McEnery}, {Michelson}, {Minuti},
  {Mitthumsiri}, {Mizuno}, {Moiseev}, {Mongelli}, {Monte}, {Monzani},
  {Moretti}, {Morselli}, {Moskalenko}, {Murgia}, {Nakajima}, {Nakamori},
  {Naumann-Godo}, {Nolan}, {Norris}, {Nuss}, {Ohno}, {Ohsugi}, {Omodei},
  {Orlando}, {Ormes}, {Ozaki}, {Paccagnella}, {Paneque}, {Panetta}, {Parent},
  {Pelassa}, {Pepe}, {Pesce-Rollins}, {Pinchera}, {Piron}, {Porter}, {Poupard},
  {Rain{\`o}}, {Rando}, {Ray}, {Razzano}, {Razzaque}, {Rea}, {Reimer},
  {Reimer}, {Reposeur}, {Ripken}, {Ritz}, {Rochester}, {Rodriguez}, {Romani},
  {Roth}, {Sadrozinski}, {Salvetti}, {Sanchez}, {Sander}, {Saz Parkinson},
  {Scargle}, {Schalk}, {Scolieri}, {Sgr{\`o}}, {Shaw}, {Siskind}, {Smith},
  {Smith}, {Spandre}, {Spinelli}, {Starck}, {Stephens}, {Striani}, {Strickman},
  {Strong}, {Suson}, {Tajima}, {Takahashi}, {Takahashi}, {Tanaka}, {Thayer},
  {Thayer}, {Thompson}, {Tibaldo}, {Tibolla}, {Tinebra}, {Torres}, {Tosti},
  {Tramacere}, {Uchiyama}, {Usher}, {Van Etten}, {Vasileiou}, {Vilchez},
  {Vitale}, {Waite}, {Wallace}, {Wang}, {Watters}, {Winer}, {Wood}, {Yang},
  {Ylinen}, {Ziegler}, \& {Fermi LAT Collaboration}}]{2010yCat..21880405A}
{Abdo}, A.~A., {Ackermann}, M., {Ajello}, M., {et~al.} 2010, \apjs, 188, 405

\bibitem[{{Abdo} {et~al.}(2011{\natexlab{a}}){Abdo}, {Ackermann}, {Ajello},
  {Antolini}, {Baldini}, {Ballet}, {Barbiellini}, {Bastieri}, {Bechtol},
  {Bellazzini}, {Berenji}, {Blandford}, {Bonamente}, {Borgland}, {Bregeon},
  {Brez}, {Brigida}, {Bruel}, {Buehler}, {Buson}, {Caliandro}, {Cameron},
  {Cannon}, {Caraveo}, {Carrigan}, {Casandjian}, {Cecchi}, {{\c{C}}elik},
  {Charles}, {Chekhtman}, {Cheung}, {Chiang}, {Ciprini}, {Claus},
  {Cohen-Tanugi}, {Conrad}, {Costamante}, {Cutini}, {Dermer}, {de Palma},
  {Donato}, {Silva}, {Drell}, {Dubois}, {Escande}, {Favuzzi}, {Fegan}, {Finke},
  {Focke}, {Fortin}, {Frailis}, {Fukazawa}, {Funk}, {Fusco}, {Gargano},
  {Gasparrini}, {Gehrels}, {Germani}, {Giglietto}, {Giordano}, {Giroletti},
  {Glanzman}, {Godfrey}, {Grenier}, {Guiriec}, {Hadasch}, {Hayashida}, {Hays},
  {Hughes}, {Itoh}, {J{\'o}hannesson}, {Johnson}, {Johnson}, {Kamae},
  {Katagiri}, {Kataoka}, {Kn{\"o}dlseder}, {Kuss}, {Lande}, {Larsson},
  {Latronico}, {Lee}, {Llena Garde}, {Longo}, {Loparco}, {Lott}, {Lovellette},
  {Lubrano}, {Makeev}, {Mazziotta}, {McEnery}, {Mehault}, {Michelson},
  {Mizuno}, {Monte}, {Monzani}, {Morselli}, {Moskalenko}, {Murgia}, {Nakamori},
  {Naumann-Godo}, {Nishino}, {Nolan}, {Norris}, {Nuss}, {Ohsugi}, {Okumura},
  {Omodei}, {Orlando}, {Ormes}, {Ozaki}, {Paneque}, {Panetta}, {Parent},
  {Pelassa}, {Pepe}, {Pesce-Rollins}, {Piron}, {Porter}, {Rain{\`o}}, {Rando},
  {Razzano}, {Reimer}, {Reimer}, {Ritz}, {Roth}, {Sadrozinski}, {Sanchez},
  {Sander}, {Schinzel}, {Sgr{\`o}}, {Siskind}, {Smith}, {Sokolovsky},
  {Spandre}, {Spinelli}, {Strickman}, {Suson}, {Takahashi}, {Tanaka}, {Thayer},
  {Thayer}, {Thompson}, {Tibaldo}, {Torres}, {Tosti}, {Tramacere}, {Uehara},
  {Usher}, {Vandenbroucke}, {Vasileiou}, {Vilchez}, {Vitale}, {Waite},
  {Wallace}, {Wang}, {Winer}, {Wood}, {Yang}, {Ylinen}, {Ziegler}, {Berdyugin},
  {Boettcher}, {Carrami{\~n}ana}, {Carrasco}, {de la Fuente}, {Diltz},
  {Hovatta}, {Kadenius}, {Kovalev}, {L{\"a}hteenm{\"a}ki}, {Lindfors},
  {Marscher}, {Nilsson}, {Pereira}, {Reinthal}, {Roustazadeh}, {Savolainen},
  {Sillanp{\"a}{\"a}}, {Takalo}, \& {Tornikoski}}]{Abd11}
{Abdo}, A.~A., {Ackermann}, M., {Ajello}, M., {et~al.} 2011{\natexlab{a}},
  \apj, 730, 101

\bibitem[{{Abdo} {et~al.}(2011{\natexlab{b}}){Abdo}, {Ackermann}, {Ajello},
  {Antolini}, {Baldini}, {Ballet}, {Barbiellini}, {Bastieri}, {Bechtol},
  {Bellazzini}, {Berenji}, {Blandford}, {Bonamente}, {Borgland}, {Bregeon},
  {Brez}, {Brigida}, {Bruel}, {Buehler}, {Buson}, {Caliandro}, {Cameron},
  {Cannon}, {Caraveo}, {Carrigan}, {Casandjian}, {Cecchi}, {{\c{C}}elik},
  {Charles}, {Chekhtman}, {Cheung}, {Chiang}, {Ciprini}, {Claus},
  {Cohen-Tanugi}, {Conrad}, {Costamante}, {Cutini}, {Dermer}, {de Palma},
  {Donato}, {Silva}, {Drell}, {Dubois}, {Escande}, {Favuzzi}, {Fegan}, {Finke},
  {Focke}, {Fortin}, {Frailis}, {Fukazawa}, {Funk}, {Fusco}, {Gargano},
  {Gasparrini}, {Gehrels}, {Germani}, {Giglietto}, {Giordano}, {Giroletti},
  {Glanzman}, {Godfrey}, {Grenier}, {Guiriec}, {Hadasch}, {Hayashida}, {Hays},
  {Hughes}, {Itoh}, {J{\'o}hannesson}, {Johnson}, {Johnson}, {Kamae},
  {Katagiri}, {Kataoka}, {Kn{\"o}dlseder}, {Kuss}, {Lande}, {Larsson},
  {Latronico}, {Lee}, {Llena Garde}, {Longo}, {Loparco}, {Lott}, {Lovellette},
  {Lubrano}, {Makeev}, {Mazziotta}, {McEnery}, {Mehault}, {Michelson},
  {Mizuno}, {Monte}, {Monzani}, {Morselli}, {Moskalenko}, {Murgia}, {Nakamori},
  {Naumann-Godo}, {Nishino}, {Nolan}, {Norris}, {Nuss}, {Ohsugi}, {Okumura},
  {Omodei}, {Orlando}, {Ormes}, {Ozaki}, {Paneque}, {Panetta}, {Parent},
  {Pelassa}, {Pepe}, {Pesce-Rollins}, {Piron}, {Porter}, {Rain{\`o}}, {Rando},
  {Razzano}, {Reimer}, {Reimer}, {Ritz}, {Roth}, {Sadrozinski}, {Sanchez},
  {Sander}, {Schinzel}, {Sgr{\`o}}, {Siskind}, {Smith}, {Sokolovsky},
  {Spandre}, {Spinelli}, {Strickman}, {Suson}, {Takahashi}, {Tanaka}, {Thayer},
  {Thayer}, {Thompson}, {Tibaldo}, {Torres}, {Tosti}, {Tramacere}, {Uehara},
  {Usher}, {Vandenbroucke}, {Vasileiou}, {Vilchez}, {Vitale}, {Waite},
  {Wallace}, {Wang}, {Winer}, {Wood}, {Yang}, {Ylinen}, {Ziegler}, {Berdyugin},
  {Boettcher}, {Carrami{\~n}ana}, {Carrasco}, {de la Fuente}, {Diltz},
  {Hovatta}, {Kadenius}, {Kovalev}, {L{\"a}hteenm{\"a}ki}, {Lindfors},
  {Marscher}, {Nilsson}, {Pereira}, {Reinthal}, {Roustazadeh}, {Savolainen},
  {Sillanp{\"a}{\"a}}, {Takalo}, \& {Tornikoski}}]{Abdo2011-II}
{Abdo}, A.~A., {Ackermann}, M., {Ajello}, M., {et~al.} 2011{\natexlab{b}},
  \apj, 730, 101

\bibitem[{{Ackermann} {et~al.}(2015){Ackermann}, {Ajello}, {Albert}, {Atwood},
  {Baldini}, {Ballet}, {Barbiellini}, {Bastieri}, {Becerra Gonzalez},
  {Bellazzini}, {Bissaldi}, {Blandford}, {Bloom}, {Bonino}, {Bottacini},
  {Bregeon}, {Bruel}, {Buehler}, {Buson}, {Caliandro}, {Cameron}, {Caputo},
  {Caragiulo}, {Caraveo}, {Cavazzuti}, {Cecchi}, {Chekhtman}, {Chiang},
  {Chiaro}, {Ciprini}, {Cohen-Tanugi}, {Conrad}, {Cutini}, {D'Ammando}, {de
  Angelis}, {de Palma}, {Desiante}, {Di Venere}, {Dom{\'\i}nguez}, {Drell},
  {Favuzzi}, {Fegan}, {Ferrara}, {Focke}, {Fuhrmann}, {Fukazawa}, {Fusco},
  {Gargano}, {Gasparrini}, {Giglietto}, {Giommi}, {Giordano}, {Giroletti},
  {Godfrey}, {Green}, {Grenier}, {Grove}, {Guiriec}, {Harding}, {Hays},
  {Hewitt}, {Hill}, {Horan}, {Jogler}, {J{\'o}hannesson}, {Johnson}, {Kamae},
  {Kuss}, {Larsson}, {Latronico}, {Li}, {Li}, {Longo}, {Loparco}, {Lott},
  {Lovellette}, {Lubrano}, {Magill}, {Maldera}, {Manfreda}, {Max-Moerbeck},
  {Mayer}, {Mazziotta}, {McEnery}, {Michelson}, {Mizuno}, {Monzani},
  {Morselli}, {Moskalenko}, {Murgia}, {Nuss}, {Ohno}, {Ohsugi}, {Ojha},
  {Omodei}, {Orlando}, {Ormes}, {Paneque}, {Pearson}, {Perkins}, {Perri},
  {Pesce-Rollins}, {Petrosian}, {Piron}, {Pivato}, {Porter}, {Rain{\`o}},
  {Rando}, {Razzano}, {Readhead}, {Reimer}, {Reimer}, {Schulz}, {Sgr{\`o}},
  {Siskind}, {Spada}, {Spandre}, {Spinelli}, {Suson}, {Takahashi}, {Thayer},
  {Thompson}, {Tibaldo}, {Torres}, {Tosti}, {Troja}, {Uchiyama}, {Vianello},
  {Wood}, {Wood}, {Zimmer}, {Berdyugin}, {Corbet}, {Hovatta}, {Lindfors},
  {Nilsson}, {Reinthal}, {Sillanp{\"a}{\"a}}, {Stamerra}, {Takalo}, \&
  {Valtonen}}]{2015ApJ...813L..41A}
{Ackermann}, M., {Ajello}, M., {Albert}, A., {et~al.} 2015, \apjl, 813, L41

\bibitem[{{Adhikari} {et~al.}(2023){Adhikari}, {Penil},
  {Westernacher-Schneider}, {Dominguez}, {Ajello}, {Buson}, {Rico}, \&
  {Zrake}}]{adh23}
{Adhikari}, S., {Penil}, P., {Westernacher-Schneider}, J.~R., {et~al.} 2023,
  arXiv e-prints, arXiv:2307.11696

\bibitem[{{Agudo} {et~al.}(2011{\natexlab{a}}){Agudo}, {Jorstad}, {Marscher},
  {Larionov}, {G{\'o}mez}, {L{\"a}hteenm{\"a}ki}, {Gurwell}, {Smith},
  {Wiesemeyer}, {Thum}, {Heidt}, {Blinov}, {D'Arcangelo}, {Hagen-Thorn},
  {Morozova}, {Nieppola}, {Roca-Sogorb}, {Schmidt}, {Taylor}, {Tornikoski}, \&
  {Troitsky}}]{Agudo2011bis}
{Agudo}, I., {Jorstad}, S.~G., {Marscher}, A.~P., {et~al.} 2011{\natexlab{a}},
  \apjl, 726, L13

\bibitem[{{Agudo} {et~al.}(2011{\natexlab{b}}){Agudo}, {Marscher}, {Jorstad},
  {Larionov}, {G{\'o}mez}, {L{\"a}hteenm{\"a}ki}, {Smith}, {Nilsson},
  {Readhead}, {Aller}, {Heidt}, {Gurwell}, {Thum}, {Wehrle}, {Nikolashvili},
  {Aller}, {Ben{\'\i}tez}, {Blinov}, {Hagen-Thorn}, {Hiriart}, {Jannuzi},
  {Joshi}, {Kimeridze}, {Kurtanidze}, {Kurtanidze}, {Lindfors}, {Molina},
  {Morozova}, {Nieppola}, {Olmstead}, {Reinthal}, {Roca-Sogorb}, {Schmidt},
  {Sigua}, {Sillanp{\"a}{\"a}}, {Takalo}, {Taylor}, {Tornikoski}, {Troitsky},
  {Zook}, \& {Wiesemeyer}}]{Agudo2011}
{Agudo}, I., {Marscher}, A.~P., {Jorstad}, S.~G., {et~al.} 2011{\natexlab{b}},
  \apjl, 735, L10

\bibitem[{{Astropy Collaboration} {et~al.}(2022){Astropy Collaboration},
  {Price-Whelan}, {Lim}, {Earl}, {Starkman}, {Bradley}, {Shupe}, {Patil},
  {Corrales}, {Brasseur}, {N{\"o}the}, {Donath}, {Tollerud}, {Morris},
  {Ginsburg}, {Vaher}, {Weaver}, {Tocknell}, {Jamieson}, {van Kerkwijk},
  {Robitaille}, {Merry}, {Bachetti}, {G{\"u}nther}, {Aldcroft},
  {Alvarado-Montes}, {Archibald}, {B{\'o}di}, {Bapat}, {Barentsen},
  {Baz{\'a}n}, {Biswas}, {Boquien}, {Burke}, {Cara}, {Cara}, {Conroy},
  {Conseil}, {Craig}, {Cross}, {Cruz}, {D'Eugenio}, {Dencheva}, {Devillepoix},
  {Dietrich}, {Eigenbrot}, {Erben}, {Ferreira}, {Foreman-Mackey}, {Fox},
  {Freij}, {Garg}, {Geda}, {Glattly}, {Gondhalekar}, {Gordon}, {Grant},
  {Greenfield}, {Groener}, {Guest}, {Gurovich}, {Handberg}, {Hart},
  {Hatfield-Dodds}, {Homeier}, {Hosseinzadeh}, {Jenness}, {Jones}, {Joseph},
  {Kalmbach}, {Karamehmetoglu}, {Ka{\l}uszy{\'n}ski}, {Kelley}, {Kern},
  {Kerzendorf}, {Koch}, {Kulumani}, {Lee}, {Ly}, {Ma}, {MacBride}, {Maljaars},
  {Muna}, {Murphy}, {Norman}, {O'Steen}, {Oman}, {Pacifici}, {Pascual},
  {Pascual-Granado}, {Patil}, {Perren}, {Pickering}, {Rastogi}, {Roulston},
  {Ryan}, {Rykoff}, {Sabater}, {Sakurikar}, {Salgado}, {Sanghi}, {Saunders},
  {Savchenko}, {Schwardt}, {Seifert-Eckert}, {Shih}, {Jain}, {Shukla}, {Sick},
  {Simpson}, {Singanamalla}, {Singer}, {Singhal}, {Sinha}, {Sip{\H{o}}cz},
  {Spitler}, {Stansby}, {Streicher}, {{\v{S}}umak}, {Swinbank}, {Taranu},
  {Tewary}, {Tremblay}, {de Val-Borro}, {Van Kooten}, {Vasovi{\'c}}, {Verma},
  {de Miranda Cardoso}, {Williams}, {Wilson}, {Winkel}, {Wood-Vasey}, {Xue},
  {Yoachim}, {Zhang}, {Zonca}, \& {Astropy Project Contributors}}]{astropy22}
{Astropy Collaboration}, {Price-Whelan}, A.~M., {Lim}, P.~L., {et~al.} 2022,
  \apj, 935, 167

\bibitem[{{Baluev}(2008)}]{2008MNRAS.385.1279B}
{Baluev}, R.~V. 2008, \mnras, 385, 1279

\bibitem[{{Bevington} \& {Robinson}(2003)}]{bev03}
{Bevington}, P.~R. \& {Robinson}, D.~K. 2003, {Data reduction and error
  analysis for the physical sciences}

\bibitem[{{Bhatta}(2017)}]{Bha17}
{Bhatta}, G. 2017, \apj, 847, 7

\bibitem[{{Breeveld} {et~al.}(2011){Breeveld}, {Landsman}, {Holland}, \&
  {et~al.}}]{brev11}
{Breeveld}, A.~A., {Landsman}, W., {Holland}, S.~T., \& {et~al.} 2011, AIPC,
  1358, 373

\bibitem[{{Burrows} {et~al.}(2005){Burrows}, {Hill}, {Nousek}, {Kennea},
  {Wells}, {Osborne}, {Abbey}, {Beardmore}, {Mukerjee}, {Short}, {Chincarini},
  {Campana}, {Citterio}, {Moretti}, {Pagani}, {Tagliaferri}, {Giommi},
  {Capalbi}, {Tamburelli}, {Angelini}, {Cusumano}, {Br{\"a}uninger}, {Burkert},
  \& {Hartner}}]{Bur05}
{Burrows}, D.~N., {Hill}, J.~E., {Nousek}, J.~A., {et~al.} 2005, \ssr, 120, 165

\bibitem[{{Chang} {et~al.}(2011){Chang}, {Ngeow}, \& {Chen}}]{Chang2011}
{Chang}, D.~C., {Ngeow}, C.~C., \& {Chen}, W.~P. 2011, in Astronomical Society
  of the Pacific Conference Series, Vol. 451, 9th Pacific Rim Conference on
  Stellar Astrophysics, ed. S.~{Qain}, K.~{Leung}, L.~{Zhu}, \& S.~{Kwok}, 143

\bibitem[{{Danforth} {et~al.}(2010){Danforth}, {Keeney}, {Stocke}, {Shull}, \&
  {Yao}}]{2010ApJ...720..976D}
{Danforth}, C.~W., {Keeney}, B.~A., {Stocke}, J.~T., {Shull}, J.~M., \& {Yao},
  Y. 2010, \apj, 720, 976

\bibitem[{{Dhiman} {et~al.}(2021){Dhiman}, {Gupta}, {Gaur}, \& {Wiita}}]{Dhi21}
{Dhiman}, V., {Gupta}, A.~C., {Gaur}, H., \& {Wiita}, P.~J. 2021, \mnras, 506,
  1198

\bibitem[{Fitzpatrick(1999)}]{Fitzpatrick1999}
Fitzpatrick, E.~L. 1999, PASP, 111, 63

\bibitem[{{Gardiner}(1994)}]{gar94}
{Gardiner}, C.~W. 1994, {Handbook of stochastic methods for physics, chemistry
  and the natural sciences}

\bibitem[{{Ghisellini} {et~al.}(1992){Ghisellini}, {Padovani}, {Celotti}, \&
  {Maraschi}}]{Ghis92}
{Ghisellini}, G., {Padovani}, P., {Celotti}, A., \& {Maraschi}, L. 1992, in
  American Institute of Physics Conference Series, Vol. 254, Testing the AGN
  paradigm, ed. S.~S. {Holt}, S.~G. {Neff}, \& C.~M. {Urry}, 398--408

\bibitem[{{Giommi} {et~al.}(2021){Giommi}, {Perri}, {Capalbi}, {D'Elia},
  {Barres de Almeida}, {Brandt}, {Pollock}, {Arneodo}, {Di Giovanni}, {Chang},
  {Civitarese}, {De Angelis}, {Leto}, {Verrecchia}, {Ricard}, {Di Pippo},
  {Middei}, {Penacchioni}, {Ruffini}, {Sahakyan}, {Israyelyan}, \&
  {Turriziani}}]{Gio21}
{Giommi}, P., {Perri}, M., {Capalbi}, M., {et~al.} 2021, \mnras, 507, 5690

\bibitem[{{Giommi} {et~al.}(2012){Giommi}, {Polenta}, {L{\"a}hteenm{\"a}ki},
  {Thompson}, {Capalbi}, {Cutini}, {Gasparrini}, {Gonz{\'a}lez-Nuevo},
  {Le{\'o}n-Tavares}, {L{\'o}pez-Caniego}, {Mazziotta}, {Monte}, {Perri},
  {Rain{\`o}}, {Tosti}, {Tramacere}, {Verrecchia}, {Aller}, {Aller},
  {Angelakis}, {Bastieri}, {Berdyugin}, {Bonaldi}, {Bonavera}, {Burigana},
  {Burrows}, {Buson}, {Cavazzuti}, {Chincarini}, {Colafrancesco}, {Costamante},
  {Cuttaia}, {D'Ammando}, {de Zotti}, {Frailis}, {Fuhrmann}, {Galeotta},
  {Gargano}, {Gehrels}, {Giglietto}, {Giordano}, {Giroletti}, {Keih{\"a}nen},
  {King}, {Krichbaum}, {Lasenby}, {Lavonen}, {Lawrence}, {Leto}, {Lindfors},
  {Mandolesi}, {Massardi}, {Max-Moerbeck}, {Michelson}, {Mingaliev}, {Natoli},
  {Nestoras}, {Nieppola}, {Nilsson}, {Partridge}, {Pavlidou}, {Pearson},
  {Procopio}, {Rachen}, {Readhead}, {Reeves}, {Reimer}, {Reinthal},
  {Ricciardi}, {Richards}, {Riquelme}, {Saarinen}, {Sajina}, {Sandri},
  {Savolainen}, {Sievers}, {Sillanp{\"a}{\"a}}, {Sotnikova}, {Stevenson},
  {Tagliaferri}, {Takalo}, {Tammi}, {Tavagnacco}, {Terenzi}, {Toffolatti},
  {Tornikoski}, {Trigilio}, {Turunen}, {Umana}, {Ungerechts}, {Villa}, {Wu},
  {Zacchei}, {Zensus}, \& {Zhou}}]{Gio2012}
{Giommi}, P., {Polenta}, G., {L{\"a}hteenm{\"a}ki}, A., {et~al.} 2012, \aap,
  541, A160

\bibitem[{{Huang} {et~al.}(2021){Huang}, {Yin}, {Hu}, {Chen}, {Jiang},
  {Alexeeva}, \& {Wang}}]{Hua21}
{Huang}, S., {Yin}, H., {Hu}, S., {et~al.} 2021, \apj, 922, 222

\bibitem[{{Kellermann}(1992)}]{Kel92}
{Kellermann}, K.~I. 1992, Science, 258, 145

\bibitem[{{Larsson}(1996)}]{lar96}
{Larsson}, S. 1996, \aaps, 117, 197

\bibitem[{{Lewis} {et~al.}(2019){Lewis}, {Finke}, \& {Becker}}]{Lew19}
{Lewis}, T.~R., {Finke}, J.~D., \& {Becker}, P.~A. 2019, \apj, 884, 116

\bibitem[{{Liodakis} \& {Petropoulou}(2020)}]{Liodakis2020-II}
{Liodakis}, I. \& {Petropoulou}, M. 2020, \apjl, 893, L20

\bibitem[{{Liodakis} {et~al.}(2018){Liodakis}, {Romani}, {Filippenko},
  {Kiehlmann}, {Max-Moerbeck}, {Readhead}, \& {Zheng}}]{Liodakis2018}
{Liodakis}, I., {Romani}, R.~W., {Filippenko}, A.~V., {et~al.} 2018, \mnras,
  480, 5517

\bibitem[{{Lomb}(1976)}]{1976Ap&SS..39..447L}
{Lomb}, N.~R. 1976, \apss, 39, 447

\bibitem[{{Maceroni} {et~al.}(2014){Maceroni}, {Lehmann}, {da Silva},
  {Montalb{\'a}n}, {Lee}, {Ak}, {Deshpande}, {Yakut}, {Debosscher}, {Guo},
  {Kim}, {Lee}, \& {Southworth}}]{Mac2014}
{Maceroni}, C., {Lehmann}, H., {da Silva}, R., {et~al.} 2014, \aap, 563, A59

\bibitem[{{Maraschi} {et~al.}(1992){Maraschi}, {Celotti}, \&
  {Ghisellini}}]{Mar92}
{Maraschi}, L., {Celotti}, A., \& {Ghisellini}, G. 1992, in Physics of Active
  Galactic Nuclei, ed. W.~J. {Duschl} \& S.~J. {Wagner}, 605

\bibitem[{{Middei} {et~al.}(2023{\natexlab{a}}){Middei}, {Liodakis}, {Perri},
  {Puccetti}, {Cavazzuti}, {Di Gesu}, {Ehlert}, {Madejski}, {Marscher},
  {Marshall}, {Muleri}, {Negro}, {Jorstad}, {Ag{\'\i}s-Gonz{\'a}lez}, {Agudo},
  {Bonnoli}, {Bernardos}, {Casanova}, {Garc{\'\i}a-Comas}, {Husillos},
  {Marchini}, {Sota}, {Kouch}, {Lindfors}, {Borman}, {Kopatskaya}, {Larionova},
  {Morozova}, {Savchenko}, {Vasilyev}, {Zhovtan}, {Casadio}, {Escudero},
  {Myserlis}, {Hales}, {Kameno}, {Kneissl}, {Messias}, {Nagai}, {Blinov},
  {Bourbah}, {Kiehlmann}, {Kontopodis}, {Mandarakas}, {Romanopoulos},
  {Skalidis}, {Vervelaki}, {Masiero}, {Mawet}, {Millar-Blanchaer},
  {Panopoulou}, {Tinyanont}, {Berdyugin}, {Kagitani}, {Kravtsov}, {Sakanoi},
  {Imazawa}, {Sasada}, {Fukazawa}, {Kawabata}, {Uemura}, {Mizuno}, {Nakaoka},
  {Akitaya}, {Gurwell}, {Rao}, {Di Lalla}, {Cibrario}, {Donnarumma}, {Kim},
  {Omodei}, {Pacciani}, {Poutanen}, {Tavecchio}, {Antonelli}, {Bachetti},
  {Baldini}, {Baumgartner}, {Bellazzini}, {Bianchi}, {Bongiorno}, {Bonino},
  {Brez}, {Bucciantini}, {Capitanio}, {Castellano}, {Ciprini}, {Costa}, {De
  Rosa}, {Del Monte}, {Di Marco}, {Doroshenko}, {Dov{\v{c}}iak}, {Enoto},
  {Evangelista}, {Fabiani}, {Ferrazzoli}, {Garcia}, {Gunji}, {Hayashida},
  {Heyl}, {Iwakiri}, {Karas}, {Kitaguchi}, {Kolodziejczak}, {Krawczynski}, {La
  Monaca}, {Latronico}, {Maldera}, {Manfreda}, {Marin}, {Marinucci}, {Massaro},
  {Matt}, {Mitsuishi}, {Ng}, {O'Dell}, {Oppedisano}, {Papitto}, {Pavlov},
  {Peirson}, {Pesce-Rollins}, {Petrucci}, {Pilia}, {Possenti}, {Ramsey},
  {Rankin}, {Ratheesh}, {Romani}, {Sgr{\'o}}, {Slane}, {Soffitta}, {Spandre},
  {Tamagawa}, {Taverna}, {Tawara}, {Tennant}, {Thomas}, {Tombesi}, {Trois},
  {Tsygankov}, {Turolla}, {Vink}, {Weisskopf}, {Wu}, {Xie}, \&
  {Zane}}]{Middei2023}
{Middei}, R., {Liodakis}, I., {Perri}, M., {et~al.} 2023{\natexlab{a}}, \apjl,
  942, L10

\bibitem[{{Middei} {et~al.}(2023{\natexlab{b}}){Middei}, {Perri}, {Puccetti},
  {Liodakis}, {Di Gesu}, {Marscher}, {Rodriguez Cavero}, {Tavecchio},
  {Donnarumma}, {Laurenti}, {Jorstad}, {Agudo}, {Marshall}, {Pacciani}, {Kim},
  {Aceituno}, {Bonnoli}, {Casanova}, {Ag{\'\i}s-Gonz{\'a}lez}, {Sota},
  {Casadio}, {Escudero}, {Myserlis}, {Sievers}, {Kouch}, {Lindfors}, {Gurwell},
  {Keating}, {Rao}, {Kang}, {Lee}, {Kim}, {Cheong}, {Jeong}, {Angelakis},
  {Kraus}, {Antonelli}, {Bachetti}, {Baldini}, {Baumgartner}, {Bellazzini},
  {Bianchi}, {Bongiorno}, {Bonino}, {Brez}, {Bucciantini}, {Capitanio},
  {Castellano}, {Cavazzuti}, {Chen}, {Ciprini}, {Costa}, {De Rosa}, {Del
  Monte}, {Di Lalla}, {Di Marco}, {Doroshenko}, {Dov{\v{c}}iak}, {Ehlert},
  {Enoto}, {Evangelista}, {Fabiani}, {Ferrazzoli}, {Garc{\'\i}a}, {Gunji},
  {Hayashida}, {Heyl}, {Iwakiri}, {Kaaret}, {Karas}, {Kislat}, {Kitaguchi},
  {Kolodziejczak}, {Krawczynski}, {La Monaca}, {Latronico}, {Maldera},
  {Manfreda}, {Marin}, {Marinucci}, {Massaro}, {Matt}, {Mitsuishi}, {Mizuno},
  {Muleri}, {Negro}, {Ng}, {O'Dell}, {Omodei}, {Oppedisano}, {Papitto},
  {Pavlov}, {Peirson}, {Pesce-Rollins}, {Petrucci}, {Pilia}, {Possenti},
  {Poutanen}, {Ramsey}, {Rankin}, {Ratheesh}, {Roberts}, {Romani}, {Sgr{\`o}},
  {Slane}, {Soffitta}, {Spandre}, {Swartz}, {Tamagawa}, {Taverna}, {Tawara},
  {Tennant}, {Thomas}, {Tombesi}, {Trois}, {Tsygankov}, {Turolla}, {Vink},
  {Weisskopf}, {Wu}, {Xie}, \& {Zane}}]{Mid23}
{Middei}, R., {Perri}, M., {Puccetti}, S., {et~al.} 2023{\natexlab{b}}, \apjl,
  953, L28

\bibitem[{{Padovani} \& {Giommi}(1995)}]{Pad95}
{Padovani}, P. \& {Giommi}, P. 1995, \mnras, 277, 1477

\bibitem[{{Pe{\~n}il} {et~al.}(2022){Pe{\~n}il}, {Ajello}, {Buson},
  {Dom{\'\i}nguez}, {Westernacher-Schneider}, \& {Zrake}}]{2022arXiv221101894P}
{Pe{\~n}il}, P., {Ajello}, M., {Buson}, S., {et~al.} 2022, arXiv e-prints,
  arXiv:2211.01894

\bibitem[{{Pe{\~n}il} {et~al.}(2024){Pe{\~n}il}, {Westernacher-Schneider},
  {Ajello}, {Dom{\'\i}nguez}, {Buson}, {Otero-Santos}, {Marcotulli},
  {Torres-Alb{\`a}}, \& {Zrake}}]{Pen23}
{Pe{\~n}il}, P., {Westernacher-Schneider}, J.~R., {Ajello}, M., {et~al.} 2024,
  \mnras, 527, 10168

\bibitem[{{Peirson} {et~al.}(2023){Peirson}, {Negro}, {Liodakis}, {Middei},
  {Kim}, {Marscher}, {Marshall}, {Pacciani}, {Romani}, {Wu}, {Di Marco}, {Di
  Lalla}, {Omodei}, {Jorstad}, {Agudo}, {Kouch}, {Lindfors}, {Aceituno},
  {Bernardos}, {Bonnoli}, {Casanova}, {Garc{\'\i}a-Comas},
  {Ag{\'\i}s-Gonz{\'a}lez}, {Husillos}, {Marchini}, {Sota}, {Casadio},
  {Escudero}, {Myserlis}, {Sievers}, {Gurwell}, {Rao}, {Imazawa}, {Sasada},
  {Fukazawa}, {Kawabata}, {Uemura}, {Mizuno}, {Nakaoka}, {Akitaya}, {Cheong},
  {Jeong}, {Kang}, {Kim}, {Lee}, {Angelakis}, {Kraus}, {Cibrario},
  {Donnarumma}, {Poutanen}, {Tavecchio}, {Antonelli}, {Bachetti}, {Baldini},
  {Baumgartner}, {Bellazzini}, {Bianchi}, {Bongiorno}, {Bonino}, {Brez},
  {Bucciantini}, {Capitanio}, {Castellano}, {Cavazzuti}, {Chen}, {Ciprini},
  {Costa}, {De Rosa}, {Del Monte}, {Di Gesu}, {Doroshenko}, {Dov{\v{c}}iak},
  {Ehlert}, {Enoto}, {Evangelista}, {Fabiani}, {Ferrazzoli}, {Garcia}, {Gunji},
  {Hayashida}, {Heyl}, {Iwakiri}, {Kaaret}, {Karas}, {Kitaguchi},
  {Kolodziejczak}, {Krawczynski}, {La Monaca}, {Latronico}, {Madejski},
  {Maldera}, {Manfreda}, {Marin}, {Marinucci}, {Massaro}, {Matt}, {Mitsuishi},
  {Muleri}, {Ng}, {O'Dell}, {Oppedisano}, {Papitto}, {Pavlov}, {Perri},
  {Pesce-Rollins}, {Petrucci}, {Pilia}, {Possenti}, {Puccetti}, {Ramsey},
  {Rankin}, {Ratheesh}, {Roberts}, {Sgr{\'o}}, {Slane}, {Soffitta}, {Spandre},
  {Swartz}, {Tamagawa}, {Taverna}, {Tawara}, {Tennant}, {Thomas}, {Tombesi},
  {Trois}, {Tsygankov}, {Turolla}, {Vink}, {Weisskopf}, {Xie}, \&
  {Zane}}]{Peirson2023}
{Peirson}, A.~L., {Negro}, M., {Liodakis}, I., {et~al.} 2023, \apjl, 948, L25

\bibitem[{{Reegen}(2007)}]{Ree07}
{Reegen}, P. 2007, \aap, 467, 1353

\bibitem[{{Roming} {et~al.}(2005){Roming}, {Kennedy}, {Mason}, {Nousek}, {Ahr},
  {Bingham}, {Broos}, {Carter}, {Hancock}, {Huckle}, {Hunsberger}, {Kawakami},
  {Killough}, {Koch}, {McLelland}, {Smith}, {Smith}, {Soto}, {Boyd},
  {Breeveld}, {Holland}, {Ivanushkina}, {Pryzby}, {Still}, \& {Stock}}]{Rom05}
{Roming}, P. W.~A., {Kennedy}, T.~E., {Mason}, K.~O., {et~al.} 2005, \ssr, 120,
  95

\bibitem[{{Scargle}(1982)}]{1982ApJ...263..835S}
{Scargle}, J.~D. 1982, \apj, 263, 835

\bibitem[{{Schlafly} \& {Finkbeiner}(2011)}]{Schlafly11}
{Schlafly}, E.~F. \& {Finkbeiner}, D.~P. 2011, ApJ, 737, 103

\bibitem[{{Sobacchi} {et~al.}(2017){Sobacchi}, {Sormani}, \&
  {Stamerra}}]{2017MNRAS.465..161S}
{Sobacchi}, E., {Sormani}, M.~C., \& {Stamerra}, A. 2017, \mnras, 465, 161

\bibitem[{{Stella} \& {Angelini}(1992)}]{Ste92}
{Stella}, L. \& {Angelini}, L. 1992, {XRONOS, a timing analysis software
  package : user's guide : version 3.00}

\bibitem[{{Sturrock} \& {Scargle}(2010)}]{Stu10}
{Sturrock}, P.~A. \& {Scargle}, J.~D. 2010, \apj, 718, 527

\bibitem[{{Tavani} {et~al.}(2018){Tavani}, {Cavaliere}, {Munar-Adrover}, \&
  {Argan}}]{2018ApJ...854...11T}
{Tavani}, M., {Cavaliere}, A., {Munar-Adrover}, P., \& {Argan}, A. 2018, \apj,
  854, 11

\bibitem[{{Urry} \& {Padovani}(1995)}]{Urr95}
{Urry}, C.~M. \& {Padovani}, P. 1995, \pasp, 107, 803

\bibitem[{{VanderPlas}(2018)}]{Van18}
{VanderPlas}, J.~T. 2018, \apjs, 236, 16

\bibitem[{{Vaughan}(2005)}]{Vau05}
{Vaughan}, S. 2005, \aap, 431, 391

\bibitem[{{Vaughan} {et~al.}(2016){Vaughan}, {Uttley}, {Markowitz},
  {Huppenkothen}, {Middleton}, {Alston}, {Scargle}, \& {Farr}}]{Vau16}
{Vaughan}, S., {Uttley}, P., {Markowitz}, A.~G., {et~al.} 2016, \mnras, 461,
  3145

\bibitem[{{Vio} {et~al.}(2013){Vio}, {Diaz-Trigo}, \& {Andreani}}]{Vio13}
{Vio}, R., {Diaz-Trigo}, M., \& {Andreani}, P. 2013, Astronomy and Computing,
  1, 5

\bibitem[{{Zdziarski} \& {Bottcher}(2015)}]{Zdz15}
{Zdziarski}, A.~A. \& {Bottcher}, M. 2015, \mnras, 450, L21

\end{thebibliography}

\end{document}